\documentclass[pdflatex,sn-mathphys-num]{sn-jnl}
\usepackage{graphicx}%
\usepackage{multirow}%
\usepackage{amsmath,amssymb,amsfonts}%
\usepackage{amsthm}%
\usepackage{mathrsfs}%
\usepackage[title]{appendix}%
\usepackage{xcolor}%
\usepackage{textcomp}%
\usepackage{manyfoot}%
\usepackage{booktabs}%
\usepackage{algorithm}%
\usepackage{algorithmicx}%
\usepackage{algpseudocode}%
\usepackage{listings}%
\usepackage{float}

\usepackage{array}%
\usepackage{longtable}  
\usepackage{ragged2e}

\usepackage{caption}      
\usepackage{subcaption}
\theoremstyle{thmstyleone}

\theoremstyle{thmstyletwo}%

\theoremstyle{thmstylethree}%

\begin{document}

\title[A systematic review of assistive technologies for children with dyslexia]{A systematic review of assistive technologies for children with dyslexia}


\author[1]{\fnm{Sansrit} \sur{Paudel}}\email{sansrit@uri.edu}

\author[1]{\fnm{Subek} \sur{Acharya}}\email{subekacharya@uri.edu}

\author[1]{\fnm{Piriyankan} \sur{Kirupaharan}}\email{pkirupaharan@uri.edu}

\author[2]{\fnm{Bishal} \sur{KC}}\email{bkc42@tntech.edu}

\author*[3]{\fnm{Bipul} \sur{Thapa}}\email{bipul@udel.edu}

\affil[1]{\orgdiv{Department of Computer Science and Information Systems}, \orgname{University of Rhode Island}, \orgaddress{
\state{Rhode Island},
\country{USA}}}

\affil[2]{\orgdiv{Department of Computer Science}, \orgname{Tennessee Technological University}, \orgaddress{
\city{Cookeville},
\state{Tennessee},
\country{USA}}}

\affil[3]{\orgdiv{Department of Computer and Information Sciences}, \orgname{University of Delaware}, \orgaddress{
\city{Newark},
\state{Delaware},
\country{USA}}}

\abstract{Dyslexia is a neurological learning disability that primarily disrupts one's ability to read, write, and spell, affecting an estimated~15-20\% of the global population. This high prevalence underscores the importance of developing effective interventions. This study presents a systematic literature review conducted between 2015 and 2024 to evaluate current trends in assistive technologies for children with dyslexia. This research shows that digital assistive technologies are leading interventions, especially with the use of mobile apps and augmented reality. More innovative technologies like virtual reality, NLP, haptic technologies, and tangible user interfaces are emerging to provide unique solutions addressing the user's needs. While non-computing devices are generally less effective in comparison to modern digital solutions, they provide a promising alternative in settings with limited access to technology.}


\maketitle

\section{Introduction}\label{sec1}


Dyslexia is a neuro-developmental disorder that affects reading, spelling, and writing abilities despite adequate intelligence and educational opportunities. Clinically, dyslexia is classified as Specific Learning Disorder with impairment in reading in the Diagnostic and Statistical Manual of Mental Disorders (DSM-5) published by the American Psychiatric Association \cite{guha2014diagnostic}. According to DSM-5, it is characterized by persistent difficulties in word reading accuracy, reading fluency, and reading comprehension that cannot be explained by intellectual disabilities, uncorrected sensory impairments, or inadequate educational instruction~\cite{guha2014diagnostic, knight2018dyslexia}.

According to international dyslexia association, dyslexia affects 15\% to 20\% of the world population\cite{ida2024}, which means that around 1 in every 5 individuals has some form of dyslexia. Dyslexia is often misunderstood as a physical condition, which is in fact a myth. The particular region of the human brain processes the smallest unit of sound called phonemes, which is a fundamental unit for identifying and decoding words. The phonemes are automatically segmented and mapped to particular words for normal individuals; however, dyslexic children struggle with this ability due to a deficit in their phonological processing capability to break down written words into phonemes. This inefficiency hinders their reading fluency and word identification despite their intact critical thinking and reasoning skills. \cite{shaywitz1996dyslexia}. Children with dyslexia typically display several key attributes, including : (a) Difficulty in comprehending multiple instructions. (b) Difficulty paying attention, sitting still, and listening to stories. (c) Difficulty with sequencing, e.g., colored beads, classroom routines. (d) Difficulty learning nursery rhymes. Hence, Dyslexic children often avoid activities that involves reading\cite{bda2024}. There are no medications or proven therapies for dyslexia, but special instruction can help people with dyslexia improve their literacy skills. The use of multi-sensory techniques helps children learn through more than one sense, which is considered an effective learning strategy for dyslexia. This technique demands visual, auditory, and kinesthetic activity \cite{saputra2018lexipal,romero2020lazy}. Various ranges of assistive technologies are prevalent to assist children with dyslexia. These include serious games, mobile apps, virtual reality, augmented reality, haptic devices, tactile tools, and non-computing devices. Apart from digital assistive technologies, which are more dominant non-computing strategies also exists however, this approach often provides ordinary teaching and learning supports that lack engagement and are considered less optimal. The rise in assistive technologies and literacy has led to the growth of technology in education. Technological advancement has opened the door to numerous possibilities for creating software products that are more interactive and can be tailored to individual needs.\\\\ This article provides the key findings on various functional elements that multiple technological and non-computing interventions offer, with the key focus on multi-sensory methodology for categorization. The primary aim of this study is to conduct a systematic literature review of assistive technologies for children with dyslexia. Unlike a narrative overview, this study follows a structured and reproducible methodology, including predefined search strategies, inclusion and exclusion criteria, and systematic study selection.
Based on this framework, the study addresses the following research questions:
\begin{itemize}
    \item RQ1: What are the current interventions that enhance the learning experience for children with dyslexia, and what are their key features?
    \item RQ2: What are the common accessibility features integrated into the existing interventions for children with dyslexia?
    \item RQ3: What gaps exist in current assistive technology research for children with dyslexia, and what opportunities exist for innovation?
\end{itemize}

The remainder of this paper is organized as follows. Section 2 provides background information on dyslexia and multi-sensory learning approaches. Section 3 describes the systematic review methodology, including the search strategy, data sources, and selection criteria. Section 4 presents the results of the review, categorized into technological and non-computational, followed by findings based on the research questions. Section 5 presents the discussion and outlines the limitations of this study. Finally, Section 6 concludes the paper.

\section{Background}
Dyslexia is a lifelong learning disorder that primarily affects reading and spelling~\cite{PETERSON20121997}. Since it affects reading, some people assume dyslexia is a problem with vision. However, it is a problem with language processing. Reading starts with being able to recognize and work with the sounds of spoken language. This skill is called phonological awareness.
Some of the common challenges faced by children with dyslexia include:
\begin{itemize}
    \item \textbf{Trouble with phonics}: Dyslexic children often don’t grasp phonics, which is a key skill for reading.

    \item \textbf{Slow to learn the alphabet}: They may struggle to remember the order of letters or associate letters with their sounds.

    \item \textbf{Difficulty with sequencing}: Tasks that require putting things in order, such as days of the week or steps in a process, are challenging.

    \item \textbf{Trouble decoding words}: They find it hard to break words into sounds (phonemes) and blend them together.

    \item \textbf{Skipping or misreading words}: They might miss small words like ``the" or confuse similar-looking words, such as ``was" and ``saw."

    \item \textbf{Poor reading comprehension}: Since reading takes so much effort, they might not understand what they’ve read, even if they can say the words.

    \item \textbf{Poor working memory}: They have difficulty holding and manipulating information in their minds, such as remembering a phone number long enough to write it down.
    \item \textbf{Easily distracted}: Because reading and writing are so challenging, they may lose focus or avoid tasks they find difficult.
    \item \textbf{Disorganized writing}: Their writing may seem jumbled, with missing words, incorrect grammar, or random capitalization because they focus so much on spelling.
\end{itemize}
Multi-sensory activities are based on whole-brain learning, which is the belief that the best way to teach concepts is by involving multiple areas in the brain\cite{park2021haptic}. Adding auditory or visual components to reading assignments, like illustrations or online activities, helps students develop stronger literacy skills\cite{NEWMAN201912}. The benefits of multi-sensory learning have been verified by contemporary research in cognitive science \cite{waterford_multisensory_learning}. Since this model has been supported by multiple researchers, one of the research goals of this study is to evaluate existing technology with respect to multi-sensory approaches (Visual, Auditory, Kinesthetic, Tactile) in system design, which fosters learning more engaging and interactive. The following contains key definitions that further elaborate the concept.
\begin{itemize}
    \item \textbf{Visual.} A visual can be anything to look at, such as a picture, diagram, or piece of film that is used to make something more appealing or easier to understand. In the human brain, dyslexia primarily affects the visual processing areas, particularly impacting the ability to rapidly identify and sequence letters on a page, often due to disruptions in the ``magnocellular pathway" which controls visual attention and eye fixation, causing difficulties with accurately perceiving the visual details of words, leading to struggles with reading fluency\cite{PROULX201416}.  
    \item \textbf{Auditory.} Audio is sound that is within the acoustic range of human hearing \cite{techtarget_audio_definition}. It is a powerful mechanism that can provide much detailed information and insights for children with dyslexia. Children with dyslexia can learn and practice a wide range of phonological sounds, distinguish similar sounds, sequencing sounds, enhancing their learning abilities.
    \item \textbf{Kinesthetic.} The term kinesthetic refers to the body movement that a learner uses to interact with their environment. The human kinesthetic system has the ability to process, integrate, and remember information, and the system is based on the immediate perception of motion impulses \cite{SHAN2021106283}.
    \item \textbf{Tactile.} The most complex and oldest-formed sensory system is the tactile sensation, which has the ability to translate information from the outside physical world into interior feelings \cite{yu2020artificial}.
\end{itemize}

\section{Methodology}
To systematically investigate current trends, accessibility features, and research gaps in assistive technologies for children with dyslexia, this study adopts a Systematic Literature Review (SLR) methodology. A SLR is a structured, transparent, and reproducible methodology designed to identify and evaluate empirical evidence~\cite{CARRERARIVERA2022101895}. This review follows the PRISMA (Preferred Reporting Items for Systematic Reviews and Meta-Analyses) framework, incorporating a well-defined search strategy with relevant keywords, explicit inclusion and exclusion criteria, and a systematic multi-stage screening process. To reflect recent technological advancements, publications spanning from 2015 to 2024 were selected for this study.

\subsection{Database source}

The literature search was conducted using three primary academic databases: ACM Digital Library, IEEE Xplore, and Google Scholar. ACM Digital Library and IEEE Xplore were selected due to their comprehensive coverage of high-quality, peer-reviewed research in computer science, where a significant proportion of assistive technology solutions are designed, implemented, and evaluated. Furthermore, given the interdisciplinary nature of dyslexia research, which spans education, cognitive science, and human–computer interaction, Google Scholar was included to broaden the search scope and capture relevant studies that may not be indexed in domain-specific engineering databases. This includes pedagogical and educational research that complements technological developments.  Mart\'{\i}n-Mart\'{\i}n et al.~\cite{martin2018google} showed that Google Scholar provides substantially broader coverage than traditional indexing services such as Web of Science and Scopus, often encompassing the majority of their indexed citations along with additional sources. 

The combination of these databases ensures a robust and comprehensive search strategy, capturing relevant studies across computing, education, and related interdisciplinary domains.

\subsection{Keyword search}
The search for literature relevant to dyslexia and children was conducted using a structured Boolean search strategy. The search string was divided into three key categories, each representing a core concept: (1) the primary topic of interest,(``dyslexia"); (2) the target demographic or context (``children" OR ``technology"); and (3) specific tools or technologies utilized in these interventions. These groups were connected using the AND operator to retrieve studies that addressed all three aspects simultaneously. To ensure comprehensive coverage, the exact search string was employed: (``dyslexia" OR "Specific learning disorder") AND (``children" OR ``technology") AND (``virtual reality" OR ``apps" OR ``mobile application" OR ``tactile" OR ``kinesthetic" OR ``auditory" OR ``games" OR ``haptic" OR ``educational games").\\
\subsection{Inclusion criteria}

To ensure a systematic and well-defined study selection process, the inclusion criteria were formulated using the PICOC (Population, Intervention, Comparison, Outcome, Context) framework. 
\begin{itemize}
    \item Population: The target demographic was strictly limited to children diagnosed with dyslexia, excluding adults and children with co-morbid conditions.
    \item Intervention: The review focused on assistive technologies, specifically incorporating multi-sensory approaches. 
    \item Comparison: The study compared the traditional intervention with digital interventions. 
    \item Outcome: The primary outcomes of interest were enhancements in learning abilities, particularly in reading, writing, spelling, and phonological awareness.
    \item Context: The literature survey was restricted to peer-reviewed, English-language publications between 2015 and 2024.
    
\end{itemize}

A set of criteria was defined to include research articles for systematic review, which are explained in Table~\ref{tab:inclusion_criteria}.

\begin{table}[h]
\caption{Inclusion Criteria}\label{tab:inclusion_criteria}
\begin{tabular*}{\textwidth}{@{\extracolsep\fill}c p{0.85\textwidth}@{}}
\toprule
\textbf{S.N} & \textbf{Inclusion Criteria} \\
\midrule
1 & Articles must be published between 2015--2024. \\
2 & Articles must be written in English. \\
3 & Publications must focus on children with dyslexia in the title or abstract. \\
4 & Any visual, auditory, kinesthetic, or tactile approach used to facilitate the learning of pupils with dyslexia in any aspect. \\
\botrule
\end{tabular*}
\end{table}
\subsection{Exclusion criteria}
Exclusion criteria exclude the irrelevant journals that were out of scope to this study, which were filtered out on the basis of a set of rules presented in Table~\ref{tab:exclusion_criteria}.
\begin{table}[h]
\caption{Exclusion Criteria}\label{tab:exclusion_criteria}
\begin{tabular*}{\textwidth}{@{\extracolsep\fill}c p{0.85\textwidth}@{}}
\toprule
\textbf{S.N} & \textbf{Criteria} \\
\midrule
1 & Articles presenting only an abstract without the full document. \\
2 & Research focus with poorly described abstract. \\
3 & Articles related to dyslexia but not relevant to children or learning populations. \\
4 & Articles focused on children with comorbid conditions. \\
5 & Age group greater than 18. \\
\botrule
\end{tabular*}
\end{table}

\subsection{Study selection}
\begin{figure}[H]
    \centering
    \includegraphics[width=0.75\textwidth]{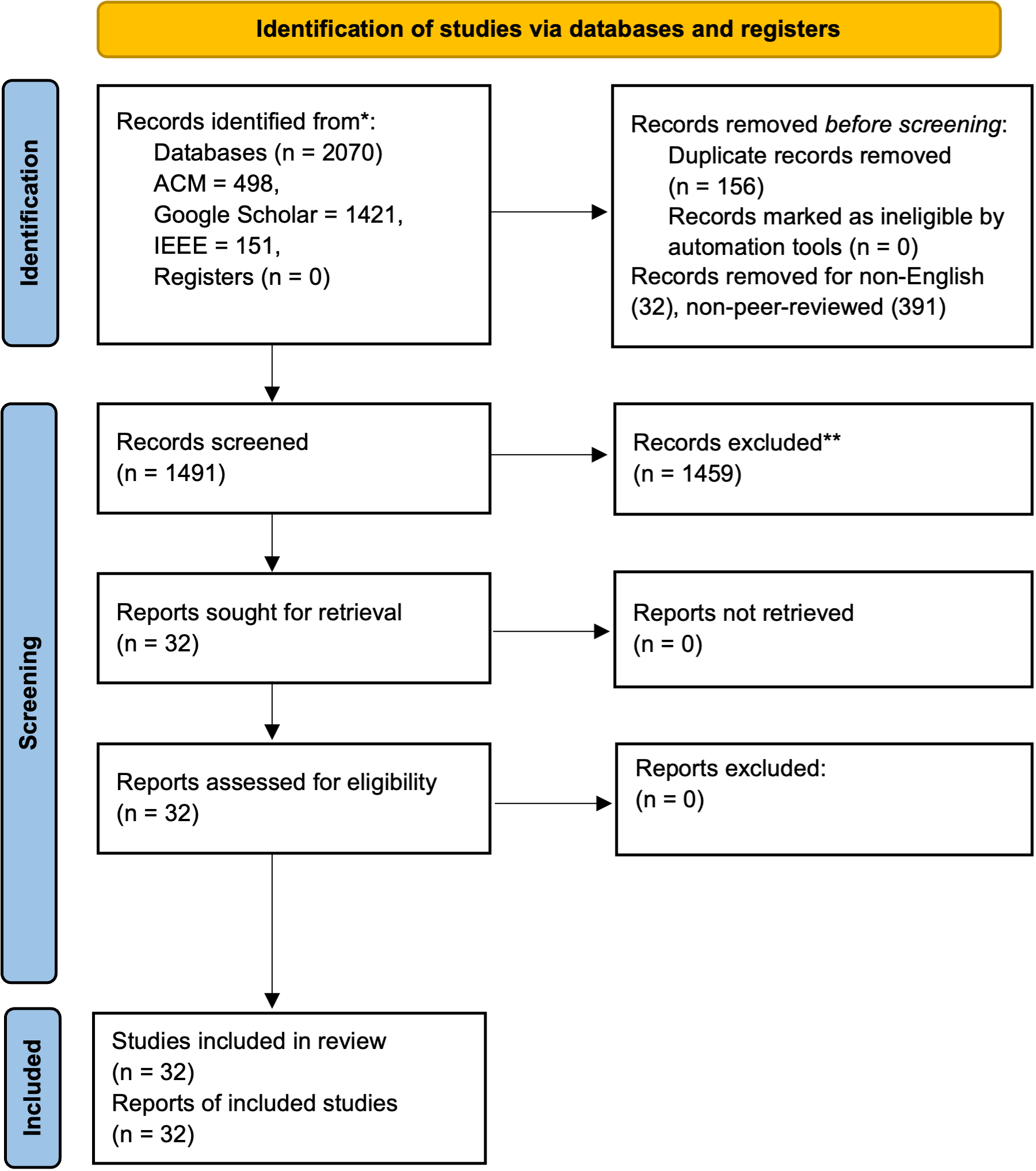}%
    \caption{
    Article selection process based on inclusion and exclusion criteria. The flow diagram shows the screening of articles from three digital libraries (ACM, Google Scholar, and IEEE), following PRISMA framework.
    }
    \label{fig:Selection flow diagram} 
\end{figure}

In order to ensure the findings directly address the research questions, inclusion and exclusion criteria were introduced. These criteria helped select relevant publications within the field and served as the initial step in the literature review. The search string was applied across major academic databases, including IEEE, Google Scholar, and ACM. Google Scholar initially returned 24200 results. From these, we selected the top 1500 most recent publications. After screening for relevance, we excluded non-pertinent papers and finalized a set of 1421 papers from this directory. In addition, the ACM digital library returned 498, and the IEEE Xplore resulted in 151 relevant papers. During the screening process, 156 duplicates were identified across the databases, 32 papers were excluded for being non-English, and 391 were excluded as they were not peer-reviewed.
The final 32 papers were selected after applying all sets of defined rules from the inclusion and exclusion criteria. The study selection process is shown in Figure~\ref{fig:Selection flow diagram}.


\section{Results}
This section presents the findings of the systematic literature review, organized into technological and non-computational interventions for children with dyslexia, followed by a synthesis of these findings based on the research questions.

\subsection{Technological Interventions}
\begin{figure}[htbp]
    \centering
    \includegraphics[width=0.7\textwidth]{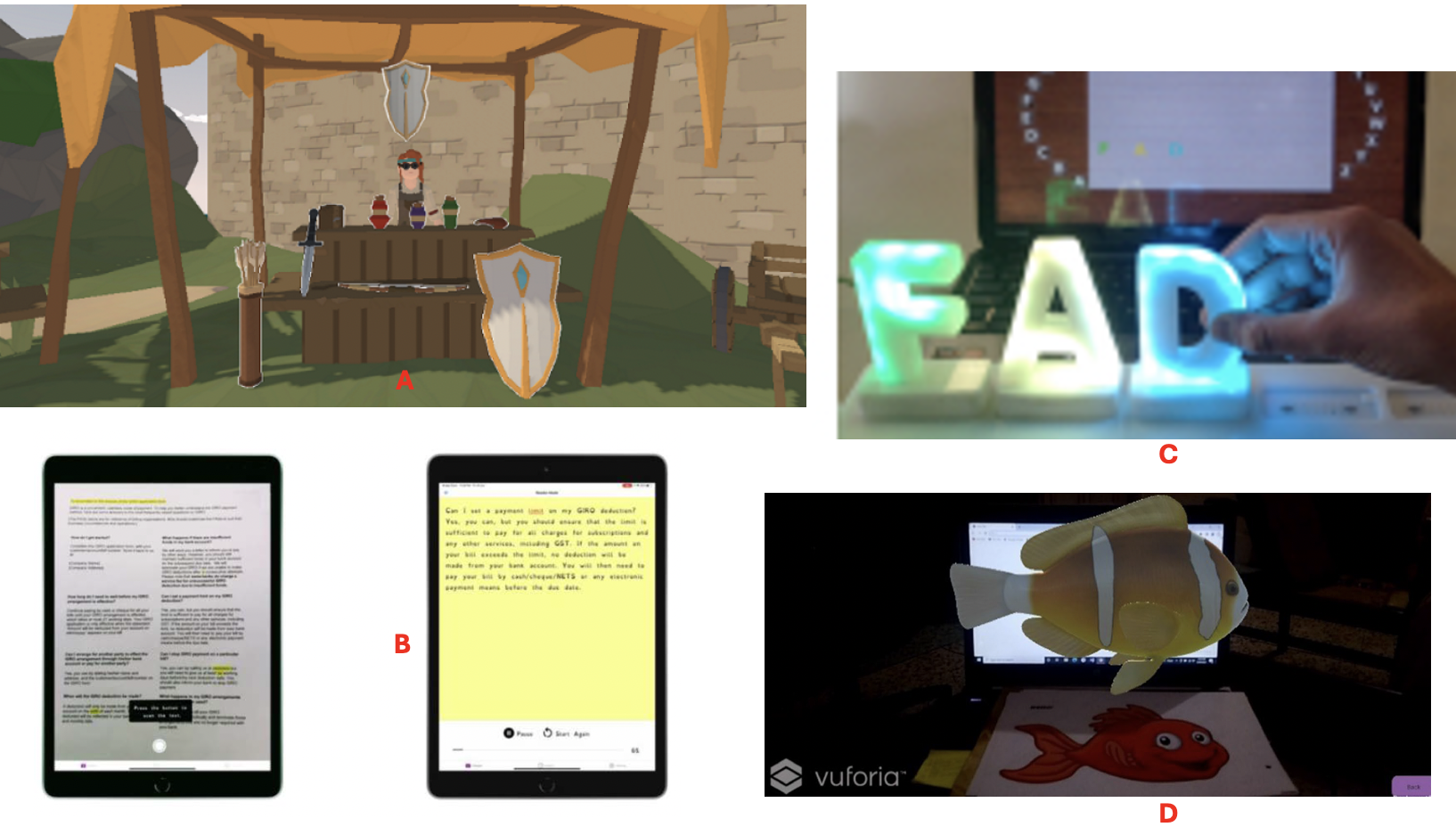} %
        \caption{Examples of assistive technologies used in interventions for dyslexia: (A) Virtual Reality environment, (B) Mobile learning applications, (C) Tangible user interface with illuminated letter blocks, and (D) Augmented Reality learning using 3D visualizations.\cite{hussain2023arlexic, gupta2021augmenta11y, saunier2022visual, polat2019tangible}}
    \label{fig:Example of assistive technologies} 
\end{figure}

Technological interventions refer to electronic, digital, or physical assistive technologies that can help people perform tasks \cite{sphero2024}. There is a wide range of assistive technologies designed for enhancing learning abilities for children with dyslexia, like augmented reality, virtual reality, mobile applications, computer applications, tangible user interface, and haptics. Many researchers conducted their research to resolve the learning disruption that children faced during the pandemic due to COVID-19. A variety of technological interventions are presented in Figure~\ref{fig:Example of assistive technologies}.
The following section will highlight key technical features that the application provides, along with their respective findings. 

\subsubsection{Mobile-Assisted Learning Applications}
Mobile-assisted learning applications are software applications designed to be used with mobile devices like smartphones or tablets~\cite{rello2012mobile}. In this literature review, mobile-assisted learning application includes all the software applications, particularly tailored for mobile and tablets, including augmented reality.

\paragraph{Augmented reality (AR)}
AR refers to the real-time integration of digital information into a user’s environment. AR technology overlays content onto the real world, enriching a user’s perception of reality rather than replacing it \cite{rahmawati2019ciselexia}. The researchers Hussain et al.~\cite{hussain2023arlexic} on designing AR games for children utilized virtual objects to enhance the reading and writing skills of children with dyslexia. The application allows children to interact with 3D alphabets and solve word puzzles, and presents an increasing level of difficulty to align with their cognitive development. This innovative approach can enhance active learning among children with dyslexia.

In this study Bhatti et al.~\cite{bhatti2020augmented} 
leverages augmented reality to create a multimedia learning application which uses AR markers or booklets to overlay 3D objects, to aid children in comprehending different educational elements such as shapes, letters, and words—tailored for children with dyslexia. Different educational practice exercises were implemented, such as recognizing sounds, understanding different animals, and their educational coursework from the specially designed educational booklet that supports AR for object identification and visualizations.

Another group of researchers Fei et al.~\cite{fei2022collectiar} 
from China has integrated augmented reality with computer vision technology to facilitate learning vocabulary and provide support for writing among children with dyslexia. This application provides children to engage in identifying real-world objects through augmented reality, where children scans real world object in their environment, and the application produces audio cues for the targeted object. CollectiAR infact, emphasizes the role of real-world engagement, encouraging children to interact with their environment to build vocabulary skills.\\
In this study Amado and Arenas et al.~\cite{lazo2023designing} 
created an augmented reality (AR) application to support children with dyslexia, focusing on usability and accessibility. Researchers adopted a user-centric design methodology to ensure the application meets the needs of its target audience. The application incorporates tasks such as word formation, letter identification, and word identification for augmented graphics, providing an engaging and interactive learning experience. While the study reported positive usability feedback, it has a limitation in that the user study was conducted with parents of dyslexic children rather than children.

\setlength{\tabcolsep}{2pt}       
\renewcommand{\arraystretch}{1.0}

\begin{longtable}{@{}
  >{\RaggedRight\arraybackslash}p{0.20\textwidth}
  >{\RaggedRight\arraybackslash}p{0.20\textwidth}
  >{\RaggedRight\arraybackslash}p{0.20\textwidth}
  >{\RaggedRight\arraybackslash}p{0.20\textwidth}
  >{\RaggedRight\arraybackslash}p{0.20\textwidth}
@{}}
\caption{Descriptive categorization of Augmented Reality applications for children with dyslexia.}
\label{tab:AR_comparison_dyslexia}\\

\toprule
\textbf{Category} & \textbf{ARLexic} & \textbf{AR Multimedia \cite{bhatti2020augmented}} & \textbf{CollectiAR} & \textbf{Educadylexia} \\
\midrule
\endfirsthead

\toprule
\textbf{Category} & \textbf{ARLexic} & \textbf{AR Multimedia \cite{bhatti2020augmented}} & \textbf{CollectiAR} & \textbf{Educadylexia} \\
\midrule
\endhead

\midrule
\multicolumn{5}{r}{\small\emph{Continued on next page}}\\
\botrule
\endfoot

\botrule
\endlastfoot

\textbf{Research Gap} &
Lack of engagement and interactivity in existing tools. &
Limited booklet design for AR. &
Existing word recognition and spelling training games were not able to teach students about their nearby environments in real time. &
Limited AR-based solutions for children with dyslexia in Latin America. \\

\textbf{Technological Features} &
Multiple gaming levels; word-scramble with timer; 3D alphabets and avatars; rewards; real-time alphabet identification for targeted physical objects; graphics and animations; feedback support. &
AR-based multimedia learning using booklets as markers to overlay 3D objects; audio cues with 3D graphics; multiple exercises to identify letters, shapes, and numbers. &
Word-hunt gameplay to find new words in the physical environment; audio cues to navigate to target objects; animated feedback; stores captured images with words for future learning; gameplay with spelling exercise; tracks objects/words collection for long-term engagement; skip control. &
Gaming levels with four activities: completing vowels; selecting correct words based on image; completing words based on image; completing complex words without images. \\

\textbf{Personalization} & Adaptable levels. & Not specified. & Search control. & Not specified. \\

\textbf{Targeted Device} & Smartphones and tablets. & Smartphones and tablets. & AR-enabled smartphones. & Android-based smartphones and tablets. \\

\textbf{Total Participants} & 21 children (ages 7–14). & Not specified; primary-level children targeted. & One child and two teachers (formative study). & 80 (parents). \\

\textbf{Validation Test} &
Paired-sample \textit{t}-test and Wilcoxon signed-rank test. &
To be evaluated. &
Formative study with qualitative feedback from teachers and children. &
Expert validation on usability, consistency, functionality, and integration; 98\% acceptance rate. \\

\textbf{Findings} &
Gamified activities such as word scramble enhance engagement. &
AR-based exercises can enhance cognitive learning. &
Potential supplementary vocabulary-learning game for training children with dyslexia. &
Cognitive game improves the learning performance of children with dyslexia. \\

\textbf{Gaming Activities or Tasks} &
Word games using 3D alphabets; puzzle games with increasing difficulty. &
Letter exercises with interactive visual and audio cues using AR booklets. &
Vocabulary scavenger hunt using computer vision to identify and spell objects in the surroundings. &
Letter recognition and comprehension tasks using AR overlays. \\

\textbf{Software Used} &
Unity 3D; Vuforia library for AR. &
Vuforia; Unity. &
Computer-vision model: SSD MobileNet V1. &
Figma; Tinkercad; MetaClassStudio. \\
\end{longtable}

An application tailored to augmented reality can further be categorized and evaluated based on multi-sensory modality, which implements a technique (visual, auditory, kinesthetic, and tactile). Since multi sensory model is considered one of the optimal models for enhancing reading abilities. This high-level categorization provides a quick reflection and implementation details to researchers on the modalities used and targeted skills. The descriptive categorization of the application is presented in Table~\ref{tab:AR_comparison_dyslexia}. 
\begin{figure}[htbp]
    \centering
    \includegraphics[width=0.7\textwidth]{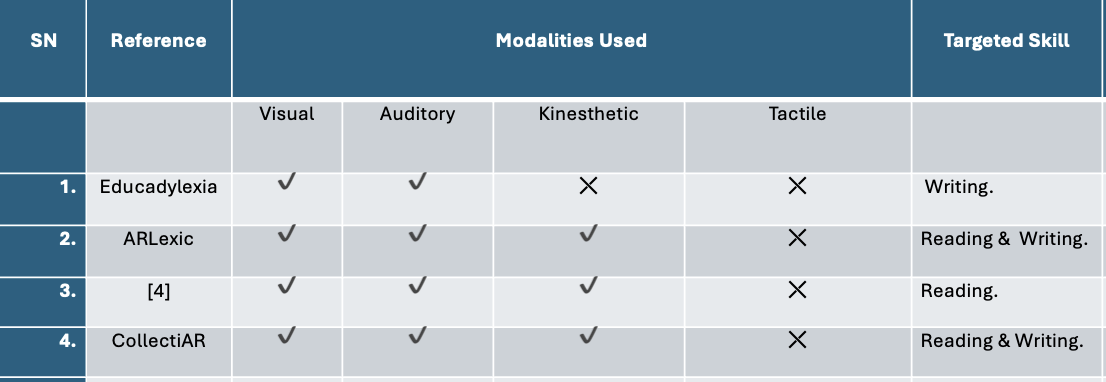} %
    \caption{Multi sensory modalities used on Augmented Reality.}
    \label{fig:Multi sensory modalities used by AR} 
\end{figure}
The Figure~\ref{fig:Multi sensory modalities used by AR} indicates that augmented reality application implements three different modalities, which include visual, auditory, and kinesthetic (users' movement), and are represented to address mixed targeted skill sets, which are crucial for enhancing the learning experience of children with dyslexia.

\paragraph{Mobile Applications}
The researchers Rello et al.~\cite{rello2013dyslexia} 
developed a mobile application designed to address the most common spelling errors and word formation for children with dyslexia. The natural language processing technique was used to process the collection of handwriting from children with dyslexia to identify the most relevant words for creating the game. Most common activities include adding, removing, and rearranging letters to create correct words. The application dynamically adjusts task difficulty based on individual performance, ensuring a tailored learning experience.

A group of researchers Vialatte et al.~\cite{vialatte2023enhancing} 
designed an application to enhance reading abilities through visual attentional training. Researchers developed symbol based visual search game where symbols most alike were presented with specific letters, making identification challenging. This exercise was proven to be efficient in enhancing visual attention and reading accuracy among children with dyslexia.

The researchers  Jaramillo-Alc´azar et al.~\cite{jaramillo2021approach} 
introduced a serious game tailored to support pattern recognition and enhance spelling abilities. In this game design, children interact with the gameplay, and their interactions are logged into a database, which is sent to a server for further evaluation that provides the therapist holistic view on analyzing the case.

The researchers Jun et al.~\cite{jun2023dysprex} 
designed game based learning system, focusing on enhancing spelling and letter identification skills. This game was designed to distinguish between most alike letters  (e.g., b/d and u/n) and improve overall reading abilities. Apart from this, the application integrates dyslexia-friendly font and design principles to ensure accessibility and ease of use.

The researchers Bigueras et al.~\cite{bigueras2020mobile} 
introduces game-based learning with a primary focus on enhancing letter and syllable recognition skills. This application provides story-based learning, which makes it unique among all other applications.

The researchers Brennan et al.~\cite{brennan2022cosmic} 
introduced a mobile game with a major focus on improving phonological awareness in children with dyslexia. This interactive game provides a wide range of activities, including exercises on developing phonological skills, such as blending of sounds, identifying the number of syllables, which are the foundational skills to excel in reading abilities. It's a co-designed application, hence, it can provide a more versatile learning experience.

The team of researchers Gupta et al.~\cite{gupta2021augmenta11y} 
introduced a reading assistant designed to support children with dyslexia through real-time text interaction. Utilizing optical character recognition (OCR) and text-to-speech (TTS) technologies, the system provides customizable dyslexia-friendly features such as adjustable fonts, colors, and word spacing. 

\begin{figure}[h]
    \centering
    \includegraphics[width=0.7\textwidth]{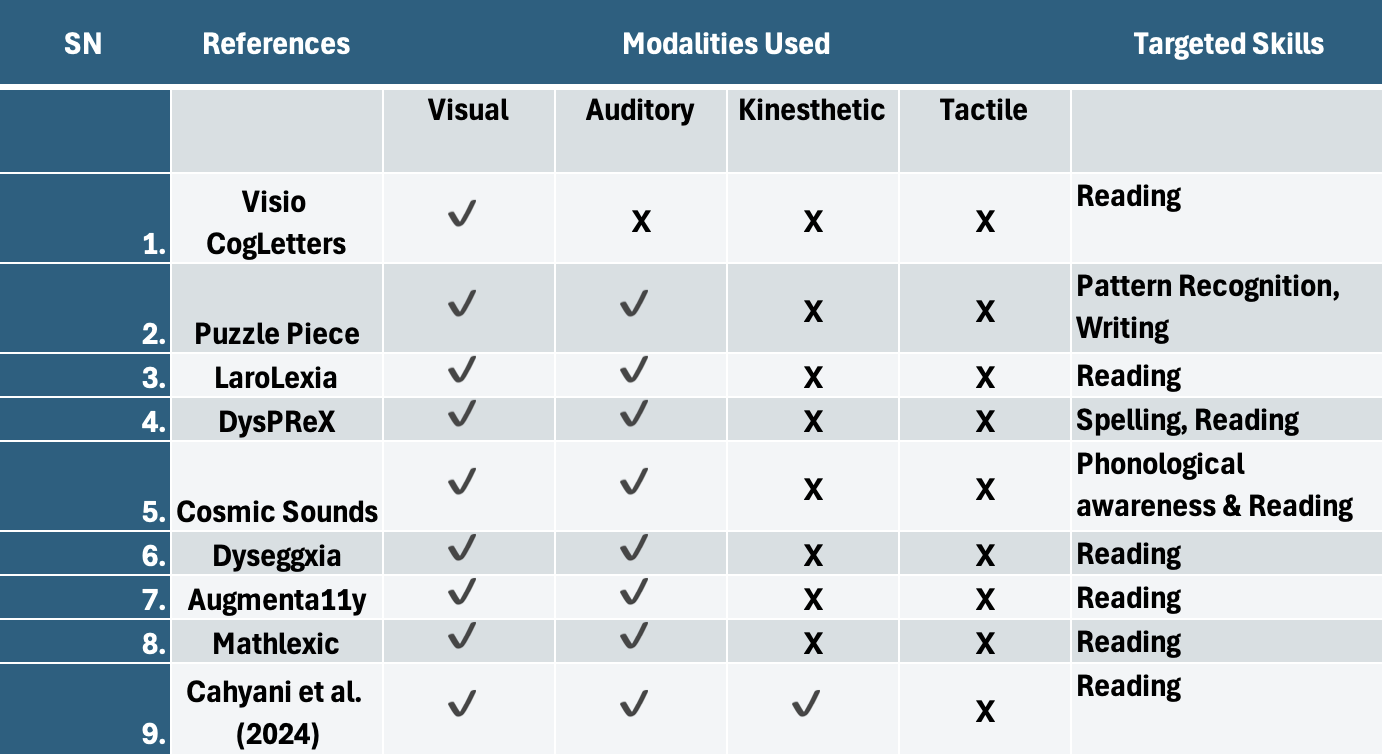} %
    \caption{Multi sensory modalities used on mobile apps.}
    \label{fig:Multisensory approach mobile apps} 
\end{figure}

The Figure~\ref{fig:Multisensory approach mobile apps} represents the multi-sensory modalities used on mobile applications, which highlights that mobile applications heavily rely on visual and auditory cues. Furthermore, mobile application addresses the demanding need for phonological awareness and reading practice among children with dyslexia. The descriptive categorization is presented in Table~\ref{tab:dyslexia_comparison_mobile_app_part1}.

\begin{table}[tbp] 
\caption{Descriptive categorization of mobile applications for children with dyslexia.}
\label{tab:dyslexia_comparison_mobile_app_part1}
\centering

\setlength{\tabcolsep}{3pt}
\renewcommand{\arraystretch}{1.12}

\begin{tabular}{@{}
  >{\RaggedRight\arraybackslash}p{0.18\linewidth}
  >{\RaggedRight\arraybackslash}p{0.205\linewidth}
  >{\RaggedRight\arraybackslash}p{0.205\linewidth}
  >{\RaggedRight\arraybackslash}p{0.205\linewidth}
  >{\RaggedRight\arraybackslash}p{0.205\linewidth}
@{}}
\toprule
\textbf{Categories} & \textbf{Dyseggxia} & \textbf{VisioCogLetters} & \textbf{Puzzle Piece} & \textbf{DysPReX} \\
\midrule
\textbf{Research Gap} & Paper-based exercises limited engagement. & Limited visual-attention training for dyslexic children. & Few cognitive games aimed at dyslexia. & Phonics-based methods are ineffective; DysPReX reduces phonics focus. \\
\textbf{Technological Features} & Six types of word exercises. & Symbol-based visual search training; challenges symbol recognition and attention. & Multiple word games; gameplay interactions logged to server. & Four gameplay types; interactive UI; audio cues; focus on visual and auditory perception. \\
\textbf{Personalization} & Adaptive exercises based on performance. & Not specified. & Selectable levels. & Customizable user settings. \\
\textbf{Targeted Device} & iOS and Android. & Not specified. & PC and mobile. & Mobile and tablets. \\
\textbf{Cross-Platform Support} & Yes. & Not specified. & No (Android only). & No (Android only). \\
\textbf{Total Participants} & 12 children with dyslexia. & 21 dyslexic children. & Ages 5--10 (count not specified). & 30 participants. \\
\textbf{Validation Methods} & Not specified. & ANOVA; Kendall's W; Wilcoxon. & Not specified. & Usability and satisfaction scores. \\
\textbf{Findings} & 5000+ downloads; engaging and adaptive vs.\ paper exercises. & Improved visual search skills. & Accessibility features + adjustable difficulty improved UX. & Positive usability (UI, functionality, learnability, satisfaction $>4/5$). \\
\textbf{Gaming Activities} & Add/remove/change letter; choose endings; cut into words. & Identify alike symbols with increasing difficulty. & Pattern recognition; spell visualized object. & Spelling and letter-differentiation (e.g., b/d, u/n). \\
\textbf{Software Used} & Not specified. & Not specified. & Unity + Firebase. & Unity 3D. \\
\botrule
\end{tabular}
\end{table}

\begin{table}[tbp]
\ContinuedFloat
\caption[]{Descriptive categorization of mobile applications for children with dyslexia (continued).}
\label{tab:dyslexia_comparison_mobile_app_part2}
\centering

\setlength{\tabcolsep}{3pt}
\renewcommand{\arraystretch}{1.12}

\begin{tabular}{@{}
  >{\RaggedRight\arraybackslash}p{0.18\linewidth}
  >{\RaggedRight\arraybackslash}p{0.205\linewidth}
  >{\RaggedRight\arraybackslash}p{0.205\linewidth}
  >{\RaggedRight\arraybackslash}p{0.205\linewidth}
  >{\RaggedRight\arraybackslash}p{0.205\linewidth}
@{}}
\toprule
\textbf{Categories} & \textbf{LaroLexia} & \textbf{Cosmic Sounds} & \textbf{Augmenta11y} & \textbf{Cahyani et al.~\cite{cahyani2024user} 
} \\
\midrule
\textbf{Research Gap} & Different learning capabilities; existing tools less efficient. & Need for phonological-awareness tools in interventions. & Limited real-time, personalized reading tools. & Limited integration of user-centered UI with gamification and multi-sensory learning. \\
\textbf{Technological Features} & Game-based reading (letter recognition, syllable reading, sight words) using familiar Filipino words. & Space-themed games for phonological awareness. & OCR-based reading assistant with customizable dyslexia-friendly settings. & Mobile UI via design thinking; guided menus, learning paths, visual/audio/kinesthetic integration, gamified tasks, adaptive progress flow. \\
\textbf{Personalization} & Tailored elements and levels. & Not specified. & Adjustable fonts, colors, spacing; TTS integration. & Gradual content unlocking by progress. \\
\textbf{Targeted Device} & Android. & Not specified. & Not specified. & Android mobile/tablet (Figma prototype). \\
\textbf{Cross-Platform Support} & No (Android only). & Not specified. & Yes (iOS and Android). & Not specified. \\
\textbf{Total Participants} & 12 children, 5 teachers, and parents. & 20 children (ages 9--12). & Not specified. & 10 dyslexic students. \\
\textbf{Validation Methods} & Pre/post paired \textit{t}-test for reading performance. & Wilcoxon signed-rank (pre/post). & Usability feedback and performance analysis. & System Usability Scale: 82. \\
\textbf{Findings} & Significant reading improvement. & Improved phonological awareness and engagement. & Increased engagement and improved reading experience. & Usable, inclusive; supports structured reading/counting practice. \\
\textbf{Gaming Activities} & Story-based, multilevel; feedback; rewards; achievements. & Interactive storyline; missions; progressive levels; rewards; supportive characters. & Scan and interact with printed text in real time. & Guessing alphabets/words; ranking modes; stepwise access to activities. \\
\textbf{Software Used} & Not specified. & Not specified. & OCR + TTS custom app. & Figma. \\
\botrule
\end{tabular}
\end{table}

\subsubsection{Tangible User Interface}
A tangible user interface (TUI) is a user interface in which a person interacts with digital information through the physical environment. The initial name was Graspable User Interface, which is no longer used. The purpose of TUI development is to empower collaboration, learning, and design by giving physical forms to digital information, thus taking advantage of the human ability to grasp and manipulate physical objects and materials\cite{10.1145/1347390.1347392}.

A group of researchers Jamali et al.~\cite{jamali2023learning} 
introduced the use of tangible user interfaces to enhance learning engagement for children with dyslexia. Tangible objects like cards and toys were integrated with a mobile application, allowing children to interact and practice phonology, spelling, and reading. The engagement and fine motor skills involved in the tangible interaction promote a deeper
understanding of sentence structure. The use of TUI with the blocks stimulates hands-on learning and facilitates the development of language skills, including grammar and syntax. The experiment with 30 children using this method resulted higher usability score, i.e. 79.5\%.

The authors Antle et al.~\cite{antle2015phonoblocks} 
designed a tactile learning system integrating 3D printed tangible letters with dynamic color cues and tactile feedback to support letter recognition and sound association for dyslexic children. The printed letters have mechanical keys that only particular letters fit on the device slots, which helps children with dyslexia to identify mirror words like 'b' and 'd'. Researchers co-designed the activities with teachers to ensure alignment with educational needs and tailored the system to promote accurate letter orientation and sound decoding. While the author had presented a pilot project, the real-world experiments are yet to be conducted; however, this system in in cooperates innovative approach to multi-sensory learning, which helps in active learning among children with dyslexia.

\begin{figure}[htbp]
    \centering
    \includegraphics[width=0.7\textwidth]{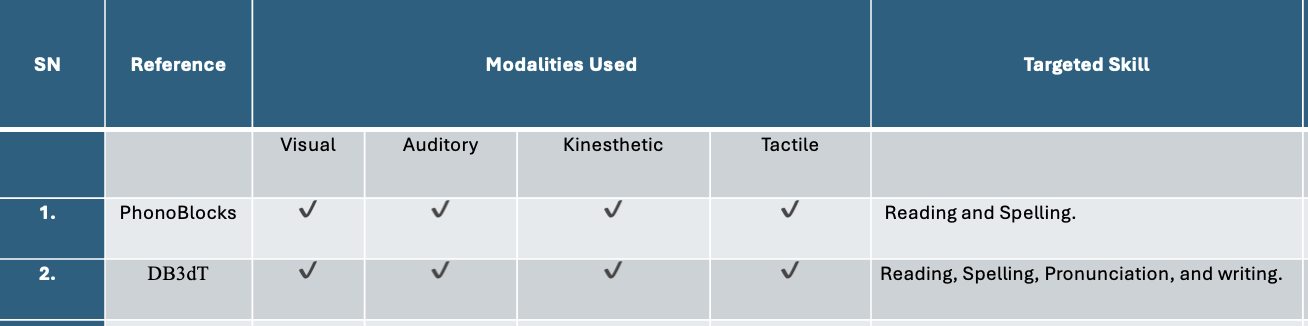} %
    \caption{Tangible user interface.}
    \label{fig:tui_table} 
\end{figure}

TUI-based interventions are summarized in Figure~\ref{fig:tui_table}, highlighting the multisensory modalities they employ (visual, auditory, kinesthetic, tactile) and the specific language skills they target, and descriptive details in Table~\ref{tab:tui_dyslexia_support}

\setlength{\tabcolsep}{4pt}
\renewcommand{\arraystretch}{1.12}

\begin{table*}[htbp]
\caption{Descriptive categorization of tangible user interfaces for children with dyslexia.}
\label{tab:tui_dyslexia_support}
\centering
\begin{tabular*}{\textwidth}{@{\extracolsep\fill}
  >{\RaggedRight\arraybackslash}p{0.18\textwidth}
  >{\RaggedRight\arraybackslash}p{0.40\textwidth}
  >{\RaggedRight\arraybackslash}p{0.40\textwidth}
@{}}
\toprule
\textbf{Category} & \textbf{DB3dT} & \textbf{PhonoBlocks} \\
\midrule
\textbf{Research Gap} &
Limited tangible and multisensory teaching methodology for children with dyslexia. &
Not specified. \\

\textbf{Technological Features} &
Tangible user interface integrated with a mobile application for enhanced learning engagement. &
Text-to-speech; colour-coded letter–sound mapping; support for mirrored letters (p, q, d, b) to reduce spelling errors; physical interaction to reinforce letter understanding. \\

\textbf{Targeted Device} &
Tablet devices. &
Touch-screen laptop with an input platform and tangible letter blocks. \\

\textbf{Targeted Skills} &
Reading, spelling, pronunciation, and writing. &
Letter recognition, sound decoding, and rectifying mirror-letter orientation. \\

\textbf{Total Participants} &
30 children with dyslexia (15 treatment, 15 control). &
No participants yet; system co-designed with teachers for future validation. \\

\textbf{Validation Methods} &
System Usability Scale (SUS); Again–Again Table; observational notes; performance checklist. &
Not specified. \\

\textbf{Findings} &
Improved engagement, learning performance, and usability scores. &
Physical letter constraints may help children use canonical letter orientation, particularly for mirrored letters (p, q, d, b). \\

\textbf{Gaming Activities} &
Vowel, syllable, and synonym exercises; interactive modules for phonology, spelling, and reading using tangible cards and toys. &
Arrange 3D-printed letters into the device slot to form words, with immediate feedback and dynamic colour changes. \\

\textbf{Software Used} &
Not specified. &
Arduino IDE for 3D-letter processing and interaction. \\
\botrule
\end{tabular*}
\end{table*}

\subsubsection{Virtual Reality}
Virtual Reality (VR) is a computer-generated environment with scenes and objects that appear to be real, making the user feel they are immersed in their surroundings. Researchers are exploring this emerging domain to investigate its potential to enhance learning abilities in children with dyslexia.

A group of researchers Pedroli et al.~\cite{pedroli2017psychometric} 
introduced a VR application for rehabilitation, particularly for children with dyslexia. The prime objective of this application was to enhance users' engagement and adherence, to address the limitations traditional therapy possesses. The application was tailored to make learning effective and enjoyable. Word identification and retaining focus in children were major activities on this system. In summary, significant improvements in word-reading test scores and homophonic writing signifying VR's potential for rehabilitation, an engaging alternative to traditional therapies.

The researchers Saunier et al.~\cite{saunier2022visual} 
experimenting with VR technology with Eye-Tracking (ET) and Brain-Computer Interface (BCI), to support children with dyslexia in therapy sessions. This prototype has not been clinically tested; however, the author's hypothesis on this technology is to monitor attention and cognitive load in real-time, thereby enabling the system to dynamically adapt exercises to the learner's needs, enhancing personalization. This is a pilot study and the research is still on early phase; however, it can introduce a novel method for combining neuroscience and VR to address the challenges faced by children with dyslexia.

A group of researchers \cite{ullah2024customizable, am2021writely} 
has developed a customizable VR Learning Environment for each basic geometric shape (cubes, cylinders, pyramids) for children with dyslexia. This application assists children in object identification and categorization tasks. Researchers had put forward multi-sensory VR realms that can bolster memory retention and comprehension through simultaneous sensory engagement. Dyslexic children can observe, listen to, and maneuver virtual entities, assisting them in grasping concepts more profoundly and cementing information more efficaciously.

The researchers Maresca et al.~\cite{maresca2022use} 
emphasizes multisensory learning in immersive VR environments. Researchers implemented user-centered designs to provide interactive reading and phonological tasks, fostering inclusion and motivation in learning processes. A significant difference in the word-reading test scores (p = 0.019) as well as in homophone writing (p = 0.034) was found. This qualitative analysis indicates higher engagement and motivation among users, presenting the effectiveness of multi-sensory VR interventions.

\begin{figure}[h]
    \centering
    \includegraphics[width=0.7\textwidth]{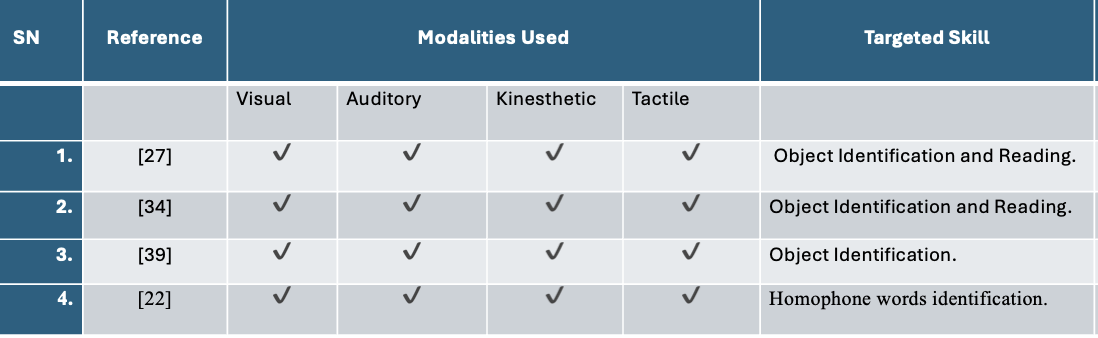} %
    \caption{Multi sensory modalities used on Virtual Reality devices.}
    \label{fig:vr_tab} 
\end{figure}

Virtual devices are used by children with dyslexia as a part of rehabilitation therapy. It is particularly used for object identification and adhering attention. The Figure~\ref{fig:vr_tab} reflects the multi-sensory modalities used by VR applications, where it encompasses all four elements with details in Table~\ref{tab:vr_dyslexia_support}

\setlength{\tabcolsep}{4pt}
\renewcommand{\arraystretch}{1.12}

\begin{table*}[htbp]
\caption{Descriptive categorization of virtual reality applications for children with dyslexia.}
\label{tab:vr_dyslexia_support}
\centering
\begin{tabular*}{\textwidth}{@{\extracolsep\fill}
  >{\RaggedRight\arraybackslash}p{0.18\textwidth}
  >{\RaggedRight\arraybackslash}p{0.205\textwidth}
  >{\RaggedRight\arraybackslash}p{0.205\textwidth}
  >{\RaggedRight\arraybackslash}p{0.205\textwidth}
  >{\RaggedRight\arraybackslash}p{0.205\textwidth}
@{}}
\toprule
\textbf{Category} &
\textbf{Pedroli et al.~\cite{pedroli2017psychometric} 
} &
\textbf{Saunier et al.~\cite{saunier2022visual}
} &
\textbf{Ullah et al.~\cite{ullah2024customizable}
} &
\textbf{Maresca et al.~\cite{maresca2022use}
} \\
\midrule

\textbf{Research Gap} &
Traditional rehabilitation techniques are repetitive, reducing engagement and adherence. &
Rehabilitation for dyslexia tends to be repetitive and discourages therapy adherence. &
Limited multisensory techniques addressing dyslexia-specific challenges. &
Limited research on VR to enhance learning ability. \\

\textbf{Technological Features} &
VR rehabilitation system with a focus on cognitive development. &
Multisensory exercises with biosensors; a hybrid VR interface combining eye tracking and brain–computer interface. &
Customizable VR environment to teach object identification and categorization. &
Immersive, multisensory VR environments with a focus on reading words. \\

\textbf{Personalization} & Not specified. & Not specified. & Not specified. & Not specified. \\

\textbf{Targeted Device} & Not specified. & Not specified. & Not specified. & Not specified. \\

\textbf{Total Participants} &
10 children with dyslexia. &
Not specified. &
Designed for children aged 6--12 (participant count not specified). &
28 children (14 control, 14 experimental). \\

\textbf{Validation Methods} &
Wechsler Intelligence Scale for Children (WISC--IV). &
Pilot test. &
Qualitative analysis and participant feedback. &
Mann--Whitney \textit{U}-test. \\

\textbf{Findings} &
Significant decrease in time to read low-frequency long words; VR devices improved attentional skills during rehabilitation. &
Multisensory VR can support rehabilitation for dyslexia; expected to improve adherence and engagement. &
Improved engagement, object-identification skills, and motivation. &
VR in rehabilitation shows promise for improving language skills, reducing disability, and promoting psychological well-being. \\

\textbf{Gaming Activities} &
Posner cueing task; attentional blink task (identify the white letter); reading words and non-words (e.g., \emph{cat}/\emph{lat}). &
Gamified 3D exercises: following/focusing on moving objects; object-finding; reading exercises. &
Interactive categorization using shapes and spatial relations. &
Object interaction in playful VR scenes; a homophone writing exercise to maintain motivation. \\

\textbf{Software Used} &
NeuroVirtual 3D. & Unity. & Not specified. & Not specified. \\
\botrule
\end{tabular*}
\end{table*}

\subsubsection{Haptics}
Haptic technology refers to anything that can create an experience of touch by applying forces, vibrations, or motions to the user \cite{sreelakshmi2017haptic}.

The researchers Park et al.~\cite{park2019investigating} 
introduced haptic technology that aids children with special abilities to improve fine motor control for their handwriting. Multilevel of haptic guidance was provided for various letters depending upon the complexity of alphabets. A group of 42 children participated in this research, and the result showed that disturbance haptic guidance was the most effective for high complexity handwriting tasks (such as writing the letters “o” and “g”), partial haptic guidance was the most effective for medium complexity handwriting tasks (such as “t,”
“r,” “s,” “e,” “n,” “a,” and “b”), and full haptic guidance was the most effective for low complexity letters (such as “i”). Haptic-based technology provides real-time feedback with visual, kinesthetic, and touch sensation which can improve brain activity in learning. The Letters that were frequently used, confusing (b and d), were prioritized to evaluate haptic systems' effectiveness in enhancing handwriting.

Similarly, Am Faleel et al.~\cite{am2021writely} 
introduced Writely, a haptic force feedback system that uses Haptic force feedback for training non-dominant hand writing. Researchers had implemented a guidance and anti-guidance system where the children who use this system would provide a forced haptic feedback if they glide the letter trace. The system is on early stage of exploration, and can be established with a low-cost solution for enhancing motor skills for special children.

\begin{figure}[htbp]
    \centering
    \includegraphics[width=0.7\textwidth]{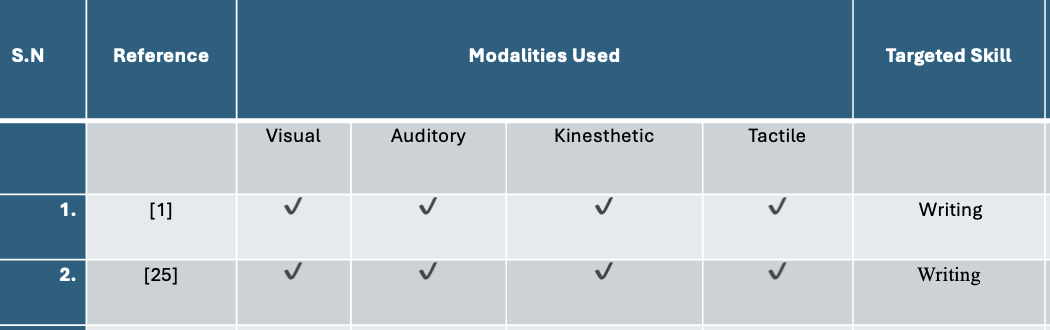} %
    \caption{Multi sensory modalities used on haptic devices.}
    \label{fig:Multisensory approach haptic} 
\end{figure}

Researchers Park et al.~\cite{park2021haptic} 
introduced falcon haptic device used in experiments. Children could grab the pencil-shaped end effector, and the haptic device guides the participants' handwriting to know how to write the task letter or draw the task shape. 12 children with cognitive and fine motor delays and mild motor difficulty participated in the study. Haptic-based learning guidance resulted to improve motor functions. Although haptic devices were not tailored to children with dyslexia, all of the technology this technology can be a handy tool for children with dyslexia to enhance their learning abilities. The haptic devices presented all modalities in enhancing writing abilities. The multi sensory model used on the haptic device is shown in Figure~\ref{fig:Multisensory approach haptic}. 

\subsection{Non-computing Interventions}
Non computing assistive interventions adopt traditional methodologies that existed before the advent of technological devices, and although primitive, this methodology offers a multi sensory approach to guiding children for their effective learning. These approaches involve physical and structured activities such as sky-writing, the use of flashcards, magnifiers, and reading guides to enhance reading and writing skills in children with dyslexia. Despite their simplicity, these methods remain effective in reinforcing foundational literacy skills and continue to be widely used, particularly in settings where access to digital technologies is limited. Researchers have implemented several non-computing strategies to enhance the learning abilities of children with dyslexia, which are briefly described below.

The authors Ziadat~\cite{ziadat2021impact} 
purposes, the VAKT modality, which includes Visual, Auditory, kinesthetic and Tactile, provides multi sensory way of learning. The key purpose of the study was to examine how effective this model is in enhancing reading and reading comprehension among students with dyslexia. Several activities were included during the experimental session such as word tracing, skywriting, and writing on the wrist. Apart from that, word cards, color pens, crayons, a mirror, sand, and a whiteboard were used to make learning more elegant and interactive. Mirror was used as a visual stimulating tool to look at himself when sounding out the phoneme, and simultaneously tapping his fingers to the thumb. Thirty-nine students participated in the survey, which revealed statistically significant improvements in comprehension reading for the experimental group that utilized the implemented multi-sensory approach (VAKT strategy).

A group of researchers Omar et al.~\cite{omar2021comparison} 
from Malaysia conducted a comparison of visual aids for enhancing reading performance in children with dyslexia. Three specific tools were evaluated: a simple magnifier, which contains a magnifying lens to enlarge text; a visual tracking magnifier, which features a guideline along with a magnifying device that is effective in reducing glare while improving readability; and lastly, a typoscope, which focuses on a line of text by blocking out extra words and glare. A case study involving 80 school children revealed that the visual tracking magnifier (VTM) intervention yielded the most significant improvement after 12 weeks of intervention. Hence, the author put forward an idea that a VTM can be used as part of a rehabilitation program for all children with dyslexia.

The authors Liu et al.~\cite{liu2021want} 
introduced a structural approach to mitigate the traditional method of teaching students with dyslexia. The practice of using the whole word to teach reading and writing for children with dyslexia has been proven inefficient. The phonics-based program with special features, including structured, cumulative, multi-sensorial activities stimulate interest, attention, as well as cultivate appropriate skills progressively. Several structural activities were designed to achieve the goal. Ranging activities from the introduction of simple vowel and consonant letters to smaller phrases and sentences were introduced. Apart from this use of story-based Fitzroy books, laminated abc cards, a word list containing a sequence of vowel, consonant alphabets, multi syllables exercise were introduced. Similarly, the physical object or materials were used in this process, which includes a set of bottle caps representing each letter (26) in the alphabet for learning the alphabet, a tactile mat to practice skywriting, white board and whiteboard markers, an exercise book, skipping ropes, hula hopes and many more. 28 total students participated in this experiment, which resulted in the experiment having improved children's progress in reading, spelling, writing, and comprehension significantly, as determined by pre-test and post-test evaluation.

A group of researcher Ferraz et al.~\cite{ferraz2018effects} 
conducted an experiment to determine the effects of a phonological remediation on reading and writing programs in children with dyslexia. Researchers evaluated phonological awareness through multiple exercise like rapid naming, working memory, reading and writing of words and non-words, and thematic writing. Each phonological intervention consists of exercises involving phonemes and syllables, syllabic and phonemic manipulation, and rhyme and alliteration. The higher complex exercise includes the addition and subtraction of
phonemes and syllables, syllabic and phonemic manipulation, oral comprehension, etc.

The authors Heikkila and Knight~\cite{heikkila2012inclusive} 
explored inclusive music teaching strategies to enhance the learning abilities of elementary-aged children with developmental dyslexia. By incorporating rhythmic, melodic, and kinesthetic activities, the study aimed to improve phonological awareness, auditory discrimination, and motor coordination. Techniques included using body movements to represent pitches, clapping rhythms, and transitioning these skills to musical instruments like the keyboard. More descriptive categorization of non computing device is illustrated by the Table~\ref{tab:dyslexia_devices_non_compute}.

\setlength{\tabcolsep}{4pt}
\renewcommand{\arraystretch}{1.12}

\begin{table*}[htbp]
\caption{Descriptive categorization of non-computing devices for children with dyslexia.}
\label{tab:dyslexia_devices_non_compute}
\centering
\begin{tabular*}{\textwidth}{@{\extracolsep\fill}
  >{\RaggedRight\arraybackslash}p{0.23\textwidth}  
  >{\RaggedRight\arraybackslash}p{0.45\textwidth}  
  >{\RaggedRight\arraybackslash}p{0.32\textwidth}  
@{}}
\toprule
\textbf{Non-computing device} & \textbf{Features} & \textbf{Benefits} \\
\midrule

\textbf{VAKT Strategy} &
Multisensory approach combining visual, auditory, kinesthetic, and tactile modalities. Activities include word tracing, skywriting, writing on the wrist, using word cards, color pens/crayons, mirror work, sand writing, and whiteboard practice. &
Improves oral reading and reading comprehension by engaging multiple senses. \\

\textbf{Visual Tracking Magnifier (VTM)} &
Magnifies text; provides a clear line guide; glare reduction via filters. &
Enhances reading speed and accuracy, reduces visual strain, and helps maintain line focus. \\

\textbf{Simple Magnifier} &
Handheld magnifier that enlarges text without additional visual aids. &
Improves readability by enlarging text for children with visual tracking challenges. \\

\textbf{Typoscope} &
Frames a single line or area of text to reduce visual distractions. &
Focuses attention on one line at a time, improving comprehension and lowering cognitive load. \\

\textbf{SMARTER Phonics Program} &
Structured phonics-based reading program. Materials/activities include story-based Fitzroy readers, laminated ABC cards, syllabified word lists, tactile mat for letter formation, and movement-based tasks (e.g., skipping ropes, hula hoops). &
Enhances decoding skills, phonemic awareness, and confidence in reading/writing—particularly helpful for rural and indigenous learners. \\

\textbf{Phonological Remediation Program} &
Targets phonological awareness, rapid naming, and reading/writing of words and non-words. Exercises include phoneme addition/deletion, syllable segmentation/manipulation, rhyme and alliteration; advanced tasks expand these operations and oral comprehension. &
Shows significant improvement in reading, writing, and memory retention through systematic intervention. \\

\textbf{Music Therapy in Education} &
Integrates melodic and rhythmic activities to build auditory discrimination and engagement. &
Improves auditory processing, phonological skills, and temporal processing, enhancing overall learning engagement. \\
\botrule
\end{tabular*}
\end{table*}

\subsection{Findings Based on Research Questions (RQ1-RQ3)}
\subsubsection{RQ1:What are the current interventions that enhance the learning experience for children with dyslexia, and what are their key features?}
The systematic literature survey resulted in finding that currently, there are two primary categories of interventions: first, being technological, and second, being non-computing. Technological intervention includes assistive technologies ranging from low-tech to high-tech devices. These include audio books, mobile apps, augmented reality, virtual reality, haptic, tangible user interface, computer apps, and natural language processing. Among all of these technological interventions, mobile application was found to be the most extensively researched domain, considering the ease of development and usability among dyslexic children. Most applications were focused on creating serious games or game-based learning interventions; however, some application like tangible user interface, were focused on providing a more realistic way of learning in-cooperating all four (VAKT) multi-sensory parameters. Apart from these, modern immersive technologies like VR and AR systems have utilized multi sensory approaches to improve letter recognition and phonological skills. Likewise, haptic-based tools have further enhanced handwriting fluency and motor coordination through tactile feedback and adaptive guidance.

Non-computing interventions were more common when adopting traditional or prevailing methods. On this method, the use of diverse activities such as the use of assistive magnifiers, typoscope, play cards, wrist writing, musical games, sand writing etc were implemented, which made learning interactive equally important in supporting dyslexic learners. Although classical, it adopts the VAKT (Visual, Auditory, Kinesthetic, Tactile) strategy to reinforce memory and improve letter recognition through touch, sound, and movement. In this model, structured word exercises were presented, which resulted on positive outcome. Furthermore, non-computing intervention presents a challenge of requiring persistent guidance from a teacher or caregiver, which makes it more dependent and less efficient compared to the use case of technological devices. 

\subsubsection{RQ2:What are the common accessibility features integrated into these interventions for children with dyslexia?}
Taking both technological and non-computing interventions into account, most of the applications targeted the common challenge of dyslexia. This includes enhancing phonological awareness and letter-sound recognition. Immersive technologies like VR and AR were focused on gamified activities that involved object and shape identification, which were useful training for adhering attention. Non-computing devices provided learning platform, engaging with physical objects like books, colors, cards, whiteboard, and they also contain structural exercise to bolster active learning. Different word games like syllabic identification, exercise related to identifying consonant and vowel sounds, spelling test, etc., were common in both technological and non-computing interventions. Many mobile applications were developed, with word games, taking into account the strategies of non-computing methodologies, thereby filling the gap it possesses. Many technological devices presented audio cues as assistive features to enhance reading fluency and accuracy; similarly, OCR and story based app were common in enhancing readability.

To sum up, gamified learning platforms were the most common features identified. Integration of rewards and points to motivate users, real-time feedback were central idea in almost every application implemented. Personalized features, including customizable applications, adjusting system settings such as font size, colors, and word spacing, making reading more comfortable, and cross-platform support, were the next big thing that some applications presented. Lastly, every application was implementing the VAKT strategy in as many possible ways to enhance the learning abilities of children with dyslexia.

\subsubsection{RQ3: What gaps exist in current assistive technology research for children with dyslexia, and what opportunities
exist for innovation?}
Current research on assistive technologies for children with dyslexia poses some gaps. First being lack of long-term studies to evaluate the efficacy of this device; there is little to no evidence on whether these benefits persist over time. The existing research was conducted with a small number of participants, in some cases even with their parents. This limits the generalizability of the system across a diverse population.

On the other side, applications integrated with a large language model or natural language processing were still limited. Their use case in children with dyslexia might result in a new finding that is yet to be explored. Most mobile applications were developed that were functional and limited to providing audio and visual cues; however, no mobile application had been designed to incorporate haptic or vibratory feedback, which could further bolster children's learning, with activated sensory feedback. All this limits the widespread adoption and effectiveness of these technologies.

There are limited applications that offer user personalization, making personalized applications the next domain to explore. Despite these gaps, there are also some opportunities for innovation in this field to consider. First, the integration of haptic feedback into mobile interfaces may enrich user interaction by providing tactile responses, thus improving engagement and accessibility. Second, the incorporation of LLMs and NLP techniques holds potential for improving the comprehension of complex sentences through context-aware simplification and semantic parsing. Lastly, the application of AI in adaptive learning environments presents a compelling direction. AI could enable real-time analysis of a child’s performance and dynamically adjust tasks to tailor their skill level, ensuring personalized and effective learning experiences.

\section{Discussion}
The systematic literature review highlights the promising role of AR in supporting children with dyslexia. AR applications, which are widely compatible with mobile and tablet devices, are particularly advantageous due to their portability, affordability, and accessibility. These applications often incorporate 3D animated graphics and interactive content, creating immersive, engaging, and multisensory learning environments that cater well to the diverse needs of dyslexic learners

Despite the advancements in assistive technologies, several limitations persist. While tools such as VR and haptic feedback systems demonstrate strong potential for enhancing learning outcomes, their high cost and technical complexity may restrict widespread adoption, especially in low-resource settings. Furthermore, VR systems can contribute to increased screen time, potentially causing eye strain or fatigue, particularly among younger users. As children undergo rapid developmental changes, their learning needs may evolve, which can render static or non-adaptive educational tools less effective over time. These factors underscore the necessity for personalized, adaptable interventions that evolve alongside the child’s cognitive and physiological growth

Similarly, applications that are tailored to learning disabilities possess varying font sizes and bright background colors, which can create a social stigma on use and can discourage them from using assistive technology, which further can limit the use case of such devices. The findings from the systematic literature directs researchers or developers on creating more personalized, cross-platform technologies incorporating multi-sensory modalities for recreating and enhancing technologies.

Another significant challenge lies in the design of these assistive technologies themselves. Tools created specifically for learners with disabilities often feature distinct interface designs—such as enlarged fonts or high-contrast colors—which, while functional, may inadvertently draw attention to the user's differences. This can contribute to social stigma, ultimately discouraging children from consistently using such tools. Therefore, ensuring that assistive technologies are discreet, inclusive, and socially acceptable is critical for long-term adoption and efficacy. The findings of our work suggest the need for more personalized, cross-platform assistive technologies that incorporate multi-sensory learning approaches.

A critical examination of the reviewed studies reveals that many are based on small sample sizes and exploratory or pilot designs, which limits the generalization of their findings. Furthermore, there is considerable variability in evaluation methods, outcome measures, and study durations, making direct comparison across studies challenging. The lack of standardized assessment frameworks also limits the ability to systematically evaluate the effectiveness of different interventions. These factors suggest that, while existing approaches are promising, the current evidence base remains preliminary and requires more rigorous, large-scale validation.

\subsection{Limitations of this study}

This study has several limitations. The review was restricted to selected databases (IEEE, ACM, and Google Scholar) and English-language publications from 2015–2024, which may introduce selection bias and exclude relevant prior or non-English work. The included studies (n = 32) are limited in number and heterogeneous in design, sample size, and evaluation methods. In addition, the findings depend on the reporting quality of primary studies, many of which involve small samples and evaluations over a short period of time.

In future work, we plan to expand database coverage, incorporate more diverse and larger-scale studies, and adopt evaluations over a long period of time to better assess the long-term effectiveness of assistive technologies for children with dyslexia.

\section{Conclusion}

The collective findings from the reviewed papers on assistive technologies for children with dyslexia highlight several key observations. Mobile-based and game-driven applications are widely used due to their accessibility and ease of use. Multi-sensory approaches are commonly applied to enhance engagement and support learning outcomes. In addition, emerging technologies such as VR, AR, haptic systems, and TUI show promising potential; however, many of these approaches are still in early stages of development and evaluation, often supported by small-scale or pilot studies.

Based on these findings, there is a need for more personalized, cross-platform assistive technologies that effectively integrate multi-sensory learning approaches. In this context, personalization refers to adapting learning experiences to individual user needs, such as adjusting difficulty levels, providing tailored feedback, or allowing customization of interface settings. Similarly, cross-platform compatibility refers to the ability of applications to function across different devices, such as mobile phones, tablets, and computers, which can improve accessibility in diverse learning environments. Multi-sensory approaches can be incorporated through combinations of visual, auditory, and interactive elements to support reading and learning processes.

Additionally, non-computing strategies continue to play a supportive role in improving phonological awareness and memory, particularly in environments with limited access to digital technologies. While existing solutions demonstrate promising directions, further research is needed to develop more robust and adaptable systems and to evaluate their effectiveness over longer periods.

\backmatter

\bmhead{Acknowledgements}

We would like to express sincere gratitude to Professor Krishna Venkatasubramanian of the University of Rhode Island for his valuable guidance, constructive feedback, and continuous support throughout the development of this work.

\section*{Author Contributions}

Sansrit Paudel led the study, including conceptualization, design of the systematic literature review methodology, literature search and study selection, data analysis, and preparation of the original manuscript draft. Subek Acharya contributed significantly to literature screening, data extraction, analysis support, and manuscript development. Piriyankan Kirupaharan and Bishal KC provided support in literature screening, data analysis, and manuscript review. Bipul Thapa contributed to conceptualization and methodology, supervised the research, validated the technical and methodological aspects, identified gaps and limitations, provided critical intellectual input, and contributed to manuscript writing, review, and editing. All authors reviewed the manuscript.

\section*{Declarations}

\subsection*{Disclosure of Interest}
The authors declare that there are no known competing interests.

\subsection*{Disclosure of funding statement}
Authors declare that they received no financial aid or funding for this work.

\subsection*{Disclosure of Data}
This study did not generate any new datasets. All data analyzed in this review were obtained from previously published studies, which are properly cited within the manuscript.

\subsection*{Ethical approval}
Not applicable.

\subsection*{Consent to Participate}
Not applicable.

\subsection*{Consent to Publish}
Not applicable.

\bibliography{bibliography}

\end{document}